\documentstyle[amsmath,amssymb,multicol,epsfig,prb,aps,preprint]{revtex}
\newcommand{\gsim}{\raisebox{-0.5ex} {$\stackrel{>} {\sim}$}}
\newcommand{\lsim}{\raisebox{-0.5ex} {$\stackrel{<} {\sim}$}}
\begin {document}

\titlepage{
\begin{center}
{\bf{\Large{Dynamic percolation theory for particle
diffusion in a polymer network}}}
\end{center}

\begin{center}
O. D\"{u}rr${}^1$, T. Volz$^{1}$, W. Dieterich${}^1$ and A. Nitzan${}^2$
\end{center}

\begin{center}
%{\it{\footnotesize{${}^1$Fakult\"{a}t f\"{u}r Physik,
%Universi\"{a}t Konstanz, D-78457 Konstanz\\
%${}^2$School of Chemistry, The Sackler Faculty of Science, Tel Aviv University,\\
%Tel Aviv 69978, Israel}}}
\small
${}^1$Fakult\"{a}t f\"{u}r Physik,
Universit\"{a}t Konstanz, D-78457 Konstanz\\
${}^2$School of Chemistry, The Sackler Faculty of Science, Tel Aviv University,\\
Tel Aviv 69978, Israel
\end{center}

\section*{Abstract}

Tracer--diffusion of small molecules through dense systems of
chain polymers is studied within an athermal lattice model, where hard
core interactions are taken into account by means of the site
exclusion principle. An approximate mapping of this problem onto
dynamic percolation theory is proposed. This method is shown to yield
quantitative results for the tracer correlation factor of the
molecules as a function of density and chain length provided the
non--Poisson character of 
temporal renewals in the disorder configurations is properly taken
into account.

\newpage
\section*{1. Introduction}

Atomic charge or mass transport processes in condensed systems often
take place in a dynamically disordered host medium, whose microscopic
structure fluctuates on a time scale of the order of the atomic
hopping time. An example of current interest in materials science are
polymer ion conductors.\cite{Gra91} These are solutions of ionic
salts in a polar polymer that can possess significant ionic
conductivities. It is well--known that ionic motions in these
materials are strongly coupled to motions of polymer chain segments, a
situation which may be viewed as implying a continuing rearrangement
of preferred ionic diffusion pathways through the host medium. At the
glass transition temperature $T_{g}$, large scale segmental motions
get frozen, suppressing long--range ionic diffusion. Other systems
where atoms diffuse in a reorganizing host medium include permeation
of small molecules through polymer films\cite{2,3} or ionic motions
through protein channels passing biological cell membranes.\cite{Hi}

Important progress in calculating the diffusion coefficient
of a random walker in a dynamically changing environment emerged from
dynamic percolation theory (DPT) and its generalizations. In its
original form due to Druger et al. \cite{Dru83,Dru85} one considers
the random walk in a bond percolation model, where configurations of
open and blocked bonds are randomly renewed at a given rate
$\lambda$. An important outcome of this model is the fact that the
frequency--dependent diffusivity $D(-i\omega,\lambda)$ can be obtained
by analytic continuation of the diffusivity
$D_{0}(-i\omega)=D(-i\omega,0)$ in the absence of renewals
\begin{equation}\label{D}
D(-i\omega,\lambda)=D_{0}(-i\omega+\lambda),
\end{equation}
irrespective of the precise form of the function $D_{0}(s)$ to be
derived from a system with only static disorder.\cite{Dru85} The same
result (\ref{D}) was independently obtained by Harrison and Zwanzig
within effective medium theory \cite{Ha85} assuming independent random
renewals of individual bonds rather than global renewals as in
Ref. \cite{Dru85}, and also by Hilfer and Orbach.\cite{Hi88} 
Subsequent work on polymer ion conductors was
focused on an identification of the central parameter of this theory,
the renewal rate $\lambda$, from experimentally observed polymer viscosities
\cite{Ni91} and more recently from dielectric relaxation
spectroscopy.\cite{Lo95} 
In parallel, the theory was generalized considerably to
off--lattice hopping \cite{Dru88}, spatially correlated renewals
\cite{Gr89}, cases with distinct kinds of migration steps
\cite{Dru91} and, in particular, to non--Poisson renewal processes
characterized by some waiting--time distribution $\psi(t)$
.\cite{Dru88} In that case the zero--frequency diffusivity in
$d=3$ dimensions is given by 
\begin{equation}\label{D1}
D=\frac{1}{6}\frac{\int_{0}^{\infty}dt\psi(t)\langle
  r^{2}(t)\rangle_{0}}{\int_{0}^{\infty}dt\,t\psi(t)},
\end{equation}
where $\langle r^{2}(t)\rangle_{0}$ is the mean--square displacement
of the random walker in a frozen environment.\cite{15} Note that in
the case $\psi(t)=\lambda\exp(-\lambda t)$ equation (\ref{D1}) reduces
to the zero--frequency limit of (\ref{D}).

While DPT or, generally, dynamic disorder hopping theory\cite{Dru88} was
developed as a framework for diffusion of small guest molecules in a fluctuating
disordered host environment, it was also recognized that the basic idea underlying these
dynamically--disordered hopping models can provide an
approximation to many--particle effects in transport processes in
interacting lattice gases.\cite{Gr90}  
A (point--like) tracer particle in an interacting lattice gas can hop to a neighbouring site provided the other particles have
arranged such that this attempted site is vacant and that energetic
conditions for the hop are fulfilled. The fact that the timescales for
the changing environment and the tracer motion are interconnected offers a way to
establish an effective dynamic bond percolation model for the tracer,
involving a time constant $\lambda$. $\lambda$ can be determined either
self--consistently or by an ansatz based on the lattice coordination number. A many--particle effective medium theory
for diffusion in interacting lattice gases emerges in this way.\cite{Gr90}

Besides these investigations for a lattice gas of point particles it
seems that DPT--theories, although motivated
by processes in polymer electrolytes, have never been tested
quantitatively in the context of statistical polymer models. While the
renewal processes associated with a system of point particles are sufficiently characterized by a single rate constant
$\lambda$ entering equation (\ref{D}),\cite{Gr90} we expect this equation to fail
for the problem of diffusion through a polymer network because of the
inherent distribution of relaxation times characterizing the
chain motion. This should result at least in a more complicated form
of the waiting--time distribution $\psi(t)$. The question now is how
far equation (\ref{D1}) can describe the diffusion coefficient, when
$\psi(t)$ is defined in a suitable way in terms of the actual dynamics
of the polymer network.

To elucidate this question, we investigate in this paper an athermal
lattice model, defined in Section 2, which consists of lattice chains
with varying density and chain length and a sufficiently dilute system
of point particles. Both chains and point particles undergo diffusion via elementary
stochastic moves. This model is a special case of a more
general lattice model of chains and point particles with specific
interactions, used previously to describe the influence of
temperature, pressure and salt--content on diffusion and network
relaxation properties of polymer electrolytes.\cite{Pe98,Du98}

In the present work, we first obtain diffusion coefficients from
dynamic Monte Carlo simulation of our model. These results serve as
a reference with respect to the subsequent approximation
method based on dynamic percolation theory. To implement this theory,
we determine by simulation i) the waiting--time
distribution $\psi(t)$, which we define in terms of the occupational
correlation function of a site next to a fixed point particle and
ii) the mean--square displacement $\langle r^{2}(t)\rangle_{0}$ of
point particles for static disorder (frozen chains). These steps are
computationally much less demanding than the full simulation. Comparison
of both methods via equation (\ref{D1}) provides a sensitive test for
the applicability of DP--theories to diffusion
in a fluctuating polymer host. We find excellent agreement between the
tracer correlation--factors as a function of density and chain length,
as obtained from those two methods. Temporal correlations,
reflected in the non--exponential character of $\psi(t)$,
are found to be crucial in this analysis.\cite{due01}

In Section 3 we specialise to a chain length $r=1$
which corresponds to a system of point--particles only, before we
present in Section 4 our full analysis for chains up to a length $r=20$. Some further
conclusions are drawn in Section 5.

\newpage

\section*{2. Simulation method and implementation of the
  dynamic percolation concept}

Consider a system of lattice chains on a three--dimensional simple
cubic lattice of spacing $a$. The chains are made of beads, assigned
to lattice sites, and linearly connected via nearest--neighbour
bonds. Apart from site exclusion, which mimics a hard--core repulsion,
no explicit interactions between beads are assumed. For $N$ chains
each with $r$ beads in a box of linear size $La$, the concentration of
occupied lattice sites is simply given by $c=Nr/L^{3}$. In addition,
our system contains point--like tracer particles, again subjected to
site exclusion, with a concentration $c_{t}\ll 1$ sufficiently small
so that correlations among them are negligible. Most of our
simulations were carried out with $L=10, r=1$ to $20$, $c_{t}=10^{-2}$, and periodic boundary conditions are employed.
After preparation of the system with the desired number of chains,
equilibration and the subsequent dynamics at equilibrium are based on
the generalized Verdier--Stockmayer
algorithm, which employs end--bond motions, kink--jumps and crankshaft
rotations \cite{Ve62,Hi75,Kr88}. Point--particles individually
perform nearest--neighbour hops. In the special case $r=2$ (moving dimers)
only the end--bond motion is active, which then is a 90--degree rotation of the
dimer about one of its end--points. As usual, introducing
\begin{equation}
D(s)=\frac{s^{2}}{6}\int_{0}^{\infty}dte^{-st}\langle r^{2}(t)\rangle
\end{equation}
we can obtain the diffusion coefficient of point--particles,
$D=\lim_{s\rightarrow 0^{+}}D(s)$, from their simulated mean--square
displacement $\langle r^{2}(t)\rangle$. To separate the average effect
of blocking, contained in a factor $1-c$, one introduces the tracer
correlation factor $f(c)\le 1$ according to 
\begin{equation}\label{D2}
D=D^{(0)}(1-c)f(c)
\end{equation}
where $D^{(0)}=\Gamma a^{2}$ denotes the diffusion coefficient for
infinite dilution $(c\rightarrow 0)$, with $\Gamma$ the bare hopping
rate.

Our aim is now to map the complete system dynamics onto a disordered
single--particle model, where disorder configurations are globally
renewed according to some appropriate waiting--time distribution $\psi(t)$. In order
to test the validity of this idea against full
simulations, we have to extract the
input quantities to equation (\ref{D1}), $\langle r^{2}(t)\rangle_{0}$ and
$\psi(t)$, from our polymer model. While $\langle r^{2}(t)\rangle_{0}$
can be obtained in a straightforward manner from separate simulations
with frozen chains, determination of $\psi(t)$ requires more
explanation. As indicated already in the Introduction, we propose to
determine $\psi(t)$ from the local occupational correlation function
$\langle n_{i}(t)n_{i}(0)\rangle$, where $n_{i}(t)$ is the occupation
by chain beads of a site $i$ adjacent to a fixed tracer
position. To simulate $\langle n_{i}(t)n_{i}(0)\rangle$ chains were
first equilibrated while keeping the tracer fixed. Such a procedure is perfectly in the spirit of dynamic
percolation theory based on renewals as ``seen'' by the tracer in its
immediate neighbourhood. Let us introduce the probability $\Phi(t)$
with $t>0$ that there is no renewal within the interval $[0,t]$ when
the foregoing renewal took place at an arbitrary time $t_{0}<0$. Following Ref. \cite{Dru88},
\begin{equation}\label{P1}
\Phi(t)=1-\int_{0}^{t}dt'\phi(t')
\end{equation}
with \cite{Fe}
\begin{equation}\label{dp}
-\frac{d\phi}{dt}=\bar{\lambda}\psi(t)
\end{equation}
where 
$\bar{\lambda}^{-1}=(\phi(0))^{-1}=\int_{0}^{\infty}dt\,t\psi(t)$
denotes the mean renewal time. Now, we argue that
with probability $\Phi(t)$ the occupation at site $i$ (next to a fixed
tracer) does not change
within $[0,t]$ so that the stochastic variable $n_{i}(t)$ (with
possible values $0$ or $1$) preserves its initial value,
$n_{i}(t)=n_{i}(0)$ and $n_{i}(t)n_{i}(0)=(n_{i}(0))^{2}=n_{i}(0)$. Conversely, with
probability $1-\Phi(t)$, one or more renewals occur within
$[0,t]$. Then, since configurations are randomly reassigned,
$n_{i}(t)$ can be replaced by its average, $c$. Hence, in this case, $n_{i}(t)n_{i}(0)=cn_{i}(0)$. Averaging in addition
over the initial occupation $n_{i}(0)$, we obtain for the correlation function
\begin{equation}\label{lni}
\langle n_{i}(t)n_{i}(0)\rangle =c[\Phi(t)+c(1-\Phi(t))].
\end{equation} 
This can be rewritten as
\begin{equation}\label{P2}
\Phi(t)=\frac{\langle n_{i}(t)n_{i}(0)\rangle -c^{2}}{c(1-c)},
\end{equation}
consistent with the requirements $\Phi(0)=1$ and $\Phi(t)\rightarrow
0$ as $t\rightarrow \infty$. Combination of (\ref{P1}), (\ref{dp}) and
(\ref{P2}) yields
\begin{eqnarray}\label{pt}
\psi(t) & = & \bar{\lambda}^{-1}\Phi''(t)\nonumber\\
& = & [\bar{\lambda}c(1-c)]^{-1}\frac{d^{2}}{dt^{2}}\langle
n_{i}(t)n_{i}(0)\rangle
\end{eqnarray}
After insertion into (\ref{D1}) the prefactors drop out. Equation
(\ref{pt}) completes the implementation of DP--theory to our many--particle model. In the next sections we test the
performance of this approximation scheme to a simple hard--core lattice
gas and to a polymeric system. 

\newpage

\section*{3. Hard--core lattice gas}

As a first application let us briefly examine the special case of
non--connected beads, $r=1$, which is identical to the conventional
hard--core lattice gas of point--particles. The tracer correlation
factor $f(c)$ in that case is known to a high degree of accuracy via
dynamic pair approximations \cite{Na80,Ta83,Di1}, giving
\begin{equation}\label{f}
f(c)=\frac{1+\langle\cos\Theta\rangle}{1-[(3-2c)/(2-c)]\langle\cos\Theta\rangle},
\end{equation}
and through simulations.\cite{Ke81} In Ref. \ref{f}, which becomes exact as
$c\rightarrow 1$, the quantity $\langle\cos\Theta\rangle$
characterizes the average directional change in two consecutive steps
of the tracer due to the presence of one vacancy. For a simple cubic
lattice, $\langle\cos\Theta\rangle\simeq -0.209$. An effective--medium
approximation to $f(c)$ was obtained recently\cite{Gr90} from dynamic percolation
theory using the Harrison Zwanzig approach.\cite{Ha85}

In what follows we apply the approach outlined in Section 2 to the
same problem. $\langle r^{2}(t)\rangle_{0}$ is obtained from
simulating a single mobile particle in the frozen configuration of the
background particles. $\Phi(t)$ is deduced from Eq. (\ref{pt}) where
$\langle n_{i}(t)n_{i}(0)\rangle$ is obtained from a short time
simulation of a lattice gas with one fixed tracer particle, as described above. These simulations
were carried out within a cubic box of length $L=10$ and periodic boundary conditions. Note that collective properties of a
hard--core lattice gas with symmetric transition rates show a
relaxational behaviour independent of concentration
\cite{Di,Ku}. The function $\Phi(t)$ as determined from (\ref{P2}) is therefore 
$c$--independent and thus can be determined from single particle
random walk theory. Within that framework $\Phi(t)$ can be interpreted
as time--dependent probability
of return of a single random walker to site $i$, taking into account
that one site adjacent to $i$ is blocked by a fixed tracer. In
Appendix A we briefly indicate  how $\Phi(t)$ can be calculated
exactly or how one can generate efficient analytic approximations.

Results for $\Phi(t)$ obtained both from Monte Carlo simulation and
from these approximations are plotted in Fig. 1. As seen from the figure, the main decay of $\Phi(t)$ at
short times is fairly well represented
by an exponential with decay rate
\begin{equation}\label{l}
\lambda_{0}=-(d\Phi/dt)_{t=0}=(5/6)\Gamma
\end{equation}
The factor $5/6$ simply arises from the fact that one of the six bonds
in the simple cubic lattice connected to site $i$ is blocked by the
tracer. The actual decay of $\Phi(t)$ is approached gradually by
continued fraction approximants of increasing order $N$, which were
derived according to Appendix A. The asymptotic decay of $\Phi(t)$ at
long times is governed by diffusion, giving $\Phi(t)\propto(\Gamma
t)^{-3/2}$, which, however, cannot be accounted for by a finite
continued fraction. For the purpose of practically evaluating
(\ref{D1}) we find that it is sufficient to approximate $\Phi(t)$ in terms of a superposition
of three exponentials. Fig. 2 shows the
$c$--dependent tracer correlation factor obtained in this way. The
agreement of data points from our DPT with the full curve representing
the dynamic pair approximation Eq. (\ref{f}) is quite
satisfactory. For completeness we also included Monte Carlo data for the full
hard core lattice gas. The DPT result for $c=1$ with value $f\simeq 0.6802$ was
obtained analytically, see Appendix A, whereas the exact value is
$f(1)=(1+\langle\cos\Theta\rangle)/(1-\langle\cos\Theta\rangle)\simeq 0.654$. Also shown are diffusion
constants calculated from the effective medium approximation as described in
Ref. \cite{Gr90}, which is based on only one time constant for renewal events. With $\lambda=\lambda_{0}$ as given by (\ref{l}) this
theory yields the dashed
curve which deviates notably from Eq. (\ref{f}) in the
high--concentration regime. (On the other hand, merely fitting $\lambda$ to the exact
value of $f(1)$,
giving $\lambda\simeq
0.62\,\Gamma$, turns out to give very good agreement
with Eq. (\ref{f}) in the whole concentration range.)

These results confirm the conclusion in Ref. \cite{Gr90} concerning the
applicability of dynamic percolation theory to many--particle systems
and at the same time indicate that the theory significantly improves
when the non--Poisson character of renewal processes is taken into
account. For the problems in the next section this last aspect will
become much more important.

\newpage

\section*{4. Lattice polymers}
For the hard-core lattice gas $(r = 1)$ the procedure described
obviously bears no computational advantage over existing methods. The
situation changes, however, when we go over to $r > 1$. 
The correlation factor $f(c)$ now depends on $r$ and no analytic
approximation equivalent to Eq. (\ref{f})
is available for this case. At the same time full simulations of the
diffusional dynamics become  more demanding because of the internal
degrees of freedom of the host molecules and the larger statistical
errors connected with the small concentration of tracer particles.
Here our approximate DBT-based computational scheme is
potentially useful. In this section we examine the performance of this
scheme. Fig. 3 summarizes our MC simulation results, again represented
in terms of the correlation factor $f(c)$, Eq. (\ref{D2}), for different chain
lengths $r$ up to $r = 20$. The full lines are fits to the simple
functional form $f(c;r)=(1-\alpha(r) c)/(1-\beta(r) c)$
with fit parameter $\alpha, \beta$ that depend on the chain lengths $r$. 
These results will be used as a basis for assessing the performance of
the approximate DP-based approach. 
As discussed above, this approach
is based on evaluating the waiting time distribution $\psi (t)$
according to Eq. (\ref{pt}) and the mean-square displacement
$\langle r^2(t) \rangle_0$ of a tracer in the presence of a frozen solvent.
Figure 4 shows typical results for the function $\Phi(t)$, see Eq.
(\ref{P2}) obtained for chains with length $r=10$ for
several concentrations. 
While our simulation results for $\langle r^2(t) \rangle_0$
are shown in Fig. 5 for the same r-values as in Fig. 4. 
Substitution of these results for  into
Eqs. (2) and (4) yields our DP-approximation for
the correlation factor $f(c)$, which is 
shown in Fig. 6 together with the 'exact' MC results of Fig. 3. 
(For clarity, only the full lines from Fig. 3, representing the fitted
data as discussed above, appear in Fig. 6). Evidently, the
DP-approximation agrees very well with the full simulation for all r.
Some further observations are noteworthy: 

(a) From the MC simulation results for $f(c;r)$ (Fig. 3) an interesting picture emerges concerning
the effect of host connectivity on the tracer diffusion. The most
prominent effect is the special behavior of chains with $r=2$ (see
below). Focusing first on chains with $r \ge 5$ we see that
for $c \lsim 0.8$ \cite{Fuss} $f(c)$ is larger than in the hard core lattice gas (Sec.
3), showing that in this density range chain connectivity facilitates
diffusion of tracer particles. For larger concentrations, however,
$f(c)$ drops markedly. 

(b) These findings for $r \gsim 5$  contrast to the behavior
found for $r=2$ which is a special
case concerning the allowed elementary moves (see Sect. 2). 
For all concentrations $c$ considered, $f(c)$ now remains larger than 0.9, 
see Fig. 3.
Dimers therefore induce only minor backward correlations in the tracer
motion. Intuitively, from the point of view of the tracer, only one
monomer of the dimer molecule effectively suppresses tracer forward motion by a
nearest-neighbor hop, while the second monomer is shielded. 

(c) The function $\Phi (t)$ is seen in Fig. 4 to
decay in a highly non-exponential fashion, indicating the importance
of temporal correlations in the associated renewal processes.
Furthermore, for small concentrations the relaxation first becomes
faster as $c$ increases, which
reflects enhanced fluctuations after the onset of overlap between
chains, whereas in the highdensity system $c = 0.8$ the decay is
markedly slowed down. For dimers ($r = 2$) we have found that $\Phi (t)$ decays
even somewhat faster than in the case $r = 1$ and is only weakly
c-dependent. 

(d) The mean square tracer displacement $\langle r^{2}(t) \rangle_0$ in a frozen
host, plotted in Fig. 5, shows a cross-over from diffusive behavior 
$\langle r^{2}(t) \rangle_0 \sim t$, to a
localized random walk 
$\langle r^{2}(t) \rangle_0 \rightarrow \text{constant as } t \rightarrow \infty$
as expected for a percolative
network. It is expected that this crossover takes place at some
critical concentration $c_{\text{crit}}$. A precise
determination of the percolation threshold $c_{\text{crit}}(r)$ for walks through a frozen network of chains of
length $r$ is beyond the scope of this article, yet rough estimates are
presented in Appendix B for 2 and 3-dimensional systems. In d = 3
dimensions, $c_{\text{crit}}$ appears to increase with r,
indicating again that for given c the frozen chains are less
prohibitive to tracer diffusion than a frozen background of
independent monomers. For example, for r = 10, the concentration
$c=0.8$ clearly exceeds $c_{\text{crit}}(r=10)$, 
(see Fig. 5 and the estimates in Appendix B).

(e) As already noted the
DP-approximation agrees very well with the full simulation for all r.
This remains true even in the special case r = 2 and has the advantage
of saving up to about one order of magnitude in computing time. }  
  
\newpage

\section*{5. Summary and Conclusion}

A method has been proposed how to map particle diffusion through a
fluctuating network of polymer chains onto dynamic percolation theory (DPT). As input quantities this theory requires the particles'
mean--square displacement $\langle r^{2}(t)\rangle_{0}$ in the frozen
network and the waiting time distribution $\psi(t)$ for network
renewals. We proposed to relate $\psi(t)$ to the occupational
correlation function of a site next to the fixed tracer particle so that
it reflects the temporal distribution of pathway openings seen by the
fixed tracer. In contrast to the standard hard core lattice
gas, $\psi(t)$ decays in a highly non--exponential fashion when longer
chains are considered. This feature of the fluctuating network appears
to be crucial in implementing dynamic percolation theory to chain systems.
When properly taken into
account, the DP--model gives quite accurate results for the tracer
correlation factor in its variation with concentration $c$ and chain
length $r$. We have verified this by comparing the results of DPT with Monte
Carlo simulations of the complete system dynamics.

Our studies so far are limited to an athermal system. The theory in
that case was found to save about one order of magnitude in computing
time relative to full simulations. Under this aspect it would be very
interesting to extend these studies by applying DP--theory with
non--Poisson renewals to interacting systems and associated questions
of dispersive transport.

\section*{Acknowledgments}
 The authors are grateful to the Lion--foundation for providing a
 travel grant. This work was also supported in part by the European
 Graduate College "Soft Condensed Matter" and by the Israel
 Science Foundation. 
\newpage

\section*{Appendix A: Hard--core lattice gas: supplementary results}

Regarding the hard--core lattice gas, the quantity $\Phi(t)$ given by
(\ref{P2}) is equal to the probability of return to site $i$ after time
$t$ in the case of a single random walker. Here site $i$ is next to a
site blocked by the tracer. In calculating $\Phi(t)$, the effect of the blocked site can
be taken into account by standard defect matrix theory.\cite{Ec} This yields
an expression for $\Phi(t)$ in terms of the solution of a $3\times 3$
matrix equation, whose coefficients are determined by unperturbed lattice Green
functions up to third--neighbor distances. The calculation is
straightforward and will not be reproduced here.

In the present context it is sufficient to obtain an accurate approximation for
$\Phi(t)$ only until it decays to about $10^{-2}$. Computationally it
is then advantageous to represent its Laplace--transform
$\hat{\Phi}(s)$ as a continued fraction of the type $\hat{\Phi}(s)=a_{0}(s+b_{1}-a_{1}(s+\ldots)^{-1})^{-1}$, generated by a short time
expansion of $\Phi(t)$.\cite{Wa73} Time derivatives $\Phi^{(n)}(t=0)$ with
$0\le n\le 2N-1$ are easily obtained by enumerating closed paths of the
walker which avoid the blocked site. 
Specifically, we use
\begin{equation}
  \label{eq:ret}
  \Phi(t) = \sum_{n=0}^{\infty} \Phi_{n} \frac{(\Gamma
  t)^{n}}{n!}e^{- \Gamma t}
\end{equation}
where $\Phi_{n}$ is the probability of return to the origin after $n$
steps.

At stage $N$ the continued
fraction is terminated such that $\hat{\Phi}(0)$ agrees with
the exact result from defect matrix theory (see above) which, at $s=0$, is determined in
terms of Watson--type integrals. Fig. 1 contains a plot of the $N$--th
order approximants for $\Phi(t)$ up to $N=6$. For $N=6$ the
simulations are accurately represented up to $t\simeq 15$.

Finally we comment on the limit $c\rightarrow 1$, where the
correlation factor from DPT can be evaluated
analytically. Obviously, the mean--square displacement $\langle
r^{2}(t)\rangle_{0}$ of a tracer in a frozen lattice is determined in
that limit by successive exchanges with one neighboring
vacancy. This gives $\langle r^{2}(t)\rangle_{0}=(1-\exp(-2\Gamma
t))/2$, where $\Gamma$ is the jump frequency. From (\ref{P2}) together
with the above--mentioned results for $\Phi(t)$ we obtain $f(1)=0.6802$ as
given in Section 3. 

\newpage

\section*{Appendix B: Percolation in a frozen network}

Calculation of the diffusion constant $D$ from (\ref{D1}) requires
knowledge of the mean square displacement $\langle r^{2}(t)\rangle_{0}$
of a tracer particle in a frozen network on time scales of the order
of the decay time of $\psi(t)$. In this Appendix we estimate
the critical concentration $c_{crit}(r)$ for percolation of a
monomer particle through a frozen network of chains, which
distinguishes diffusive from localised behavior of $\langle
r^{2}(t)\rangle_{0}$ at long times. To our knowledge, this problem of
correlated percolation has not been investigated before for general
chainlength $r$. We do not, however, attempt any precise determination
of $c_{crit}(r)$; rather we like to point out some major qualitative
trends in $c_{crit}(r)$ as a function of $r$ in $d=3$ and $d=2$ dimensions. 

In $d=3$, equilibrated chain configurations were prepared by the same
algorithm as described in Section 2, while in $d=2$ we used the
algorithm by Siepmann et al.\cite{Bi} For systems of varying size $L$,
increased in steps $\Delta L$, we
determined the probabilities $P(r,c,L)$ of occurrence of a spanning cluster of
vacant nearest--neighbor sites. For fixed $r$, we obtained points
of intersection of successive curves $P(r,c,L)$ and $P(r,c,L-\Delta
L)$ versus $c$, which give successive approximations for
$c_{crit}(r)$.\cite{Staufer} Estimates for $c_{crit}(r)$ were deduced from
calculations with $L$--values up to a maximum $L_{max}$, to be adapted to $r$. For example, in $d=2$ we went up
to $L_{max}=200$ for $r=20$, whereas $L_{max}=80$ for frozen dimers $r=2$.
These values appeared sufficient to achieve reasonable convergence. In the limit
$r=1$ we find $c_{crit}\simeq 0.41$ for $d=2$ and $c_{crit}=0.69$ for
$d=3$, which reasonably agree with values $1-p_{c}$ given by the
well--known thresholds $p_{c}$ for site percolation on square and
simple cubic lattices, respectively. The qualitative $r$--dependence of
$c_{crit}(r)$ is shown in Fig. 7. In $d=3$, $c_{crit}(r)$
monotonously increases with $r$ which we interpret as a reduction of
blocking of open pathways through the connectivity of chains. The most
pronounced increase occurs already when going from monomers to
$r=2$. In $d=2$ this argument
again applies to the step from $r=1$ to $r=2$, but blocking becomes
more effective for longer chains so that $c_{crit}(r)$
decreases, with $c_{crit}(r)<c_{crit}(1)$ for $r\gsim 12$. For long chains, especially
in $d=2$, it might be interesting to study $\langle
r^{2}(t)\rangle_{0}$ near criticality on different length scales above
and below the radius of gyration of chains, but this goes beyond the scope of
this work.

Generally, for sufficiently long times and small
$|c-c_{crit}|$ such that the correlation length becomes larger than
the size of chains, one expects $\langle r^{2}(t)\rangle_{0}$ to
follow the standard scaling forms for non--correlated percolation,
which imply associated scaling forms for the frequency--dependent
diffusivity $D_{0}(-i\omega)$ as $\omega\rightarrow 0$. In
applications of DP--theory where the
analytic continuation rule (\ref{D}) holds at least for small $\lambda$, one can immediately predict the
asymptotic forms of the long--time diffusion constant $D=D_{0}(\lambda)$ of a
walker in the presence of slow ($\lambda\rightarrow 0$), Poisson--type
network renewals.\cite{Di99} The resulting scaling expressions for $D$ are
straightforward to write down from the corresponding expressions for
$D_{0}(-i\omega)$ given in Ref. \cite{Re}. Some
notable special cases for nonzero $\lambda$ are
\begin{equation}
D\sim\left\{\begin{array}{cc}\lambda^{1-k};&c=c_{crit};\\\lambda|c-c_{crit}|^{2\nu-\beta};&c>c_{crit}\end{array}\right.
\end{equation}
where $\nu$ and $\beta$ are the conventional static percolation
exponents for the correlation length
and the order parameter, respectively, and $k$ is the dynamic
critical exponent for anomalous diffusion in a percolation system at
criticality.

\newpage 

\section*{Figure captions}

\begin{itemize}
\item[Fig. 1] Semilogarithmic plot of the function $\Phi(t)$ (see
  equation 8) for the hard core lattice gas. Monte Carlo data are
  obtained by simulating the probability of return of a single random
  walker starting next to a blocked site.

Dashed-dotted line: single exponential approximation as determined by the
initial slope (see equation 11). The other thin lines represent
continued fraction approximants up to order $N=6$ (see Appendix A).
\item[Fig. 2] Tracer correlation factor $f(c)$ of the hard core lattice gas
  against concentration, obtained by different methods (see text).
\item[Fig. 3] Simulated tracer correlation factor $f(c)$ for different chain lengths
  $r=2,5,10$ and $20$. Full lines refer to the fit function $f
  \approx (1-\alpha c) / (1-\beta c)$, where for $r=5,10$ and $20$ the
  fit parameters are $\beta=1$ and 
   $\alpha=1.057,1.062$ and $1.071$, respectively. 
   This implies that the diffusion constant (\ref{D2}) is
   approximately linear in $c$, $D(c) \simeq D_{0}(1-\alpha c)$.
   One the other hand, for $r=2$ we find $\alpha =0.391$ and
   $\beta=0.318.$
   For comparison we also show simulation data for $r=1$ (hard core
   lattice gas) together with Eq. (\ref{f}) (dashed-dotted line).
\item[Fig. 4] $\Phi(t)$ for chains of length $r=10$ for three
  different concentrations $c=0.8,0.1$ and $0.4$ (from above). Also shown is
the short time behavior of $\Phi(t)$ for $r=1$ (dashed-dotted line), reproduced from Fig. 1.
\item[Fig. 5] Mean--square displacement $\langle r^{2}(t)\rangle_{0}$
  of walkers in a frozen chain network. Parameters are as in
  Fig. 4. For $c=0.4$ a comparison is made with the case $r=1$.
\item[Fig. 6] Comparison of tracer correlation factors from DP--theory
  for chains of different lengths (data points) with results from full
  simulations. Full lines represent fit functions for the simulation
  data, reproduced from Fig. 3.
\item[Fig. 7] Estimates for the critical concentration $c_{crit}(r)$
  versus $r$ for percolation of a monomer particle through a frozen
  network of chains of length $r$ on simple cubic ($d=3$) and a square
  ($d=2$) lattices.
\end{itemize}

\newpage 

\begin{figure}[htb]
  \begin{center}
   \epsfig{file=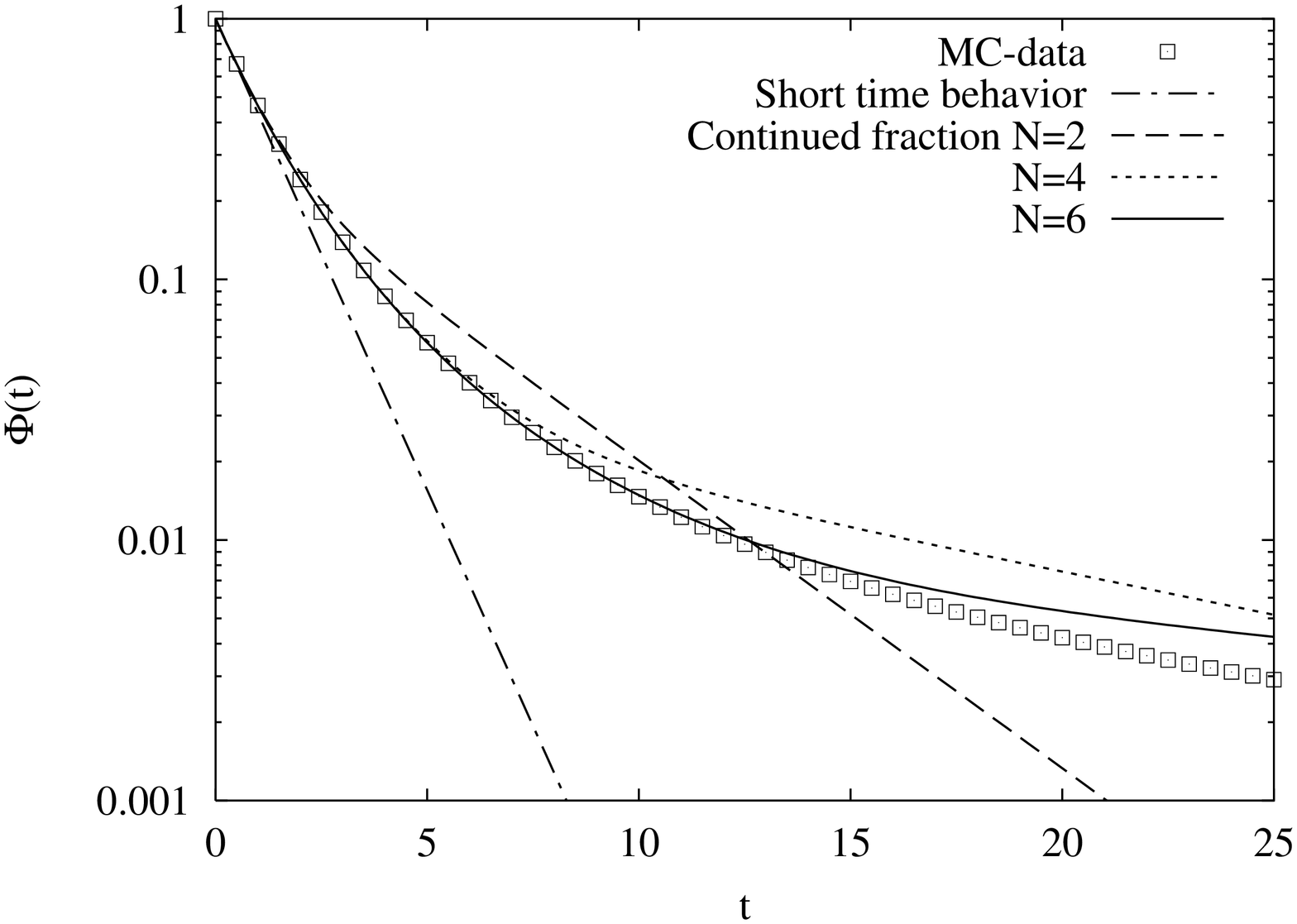,width=0.75\linewidth,angle=-90}
   Fig 1.
 \end{center}
\end{figure}
\begin{figure}[htb]
  \begin{center}
   \epsfig{file=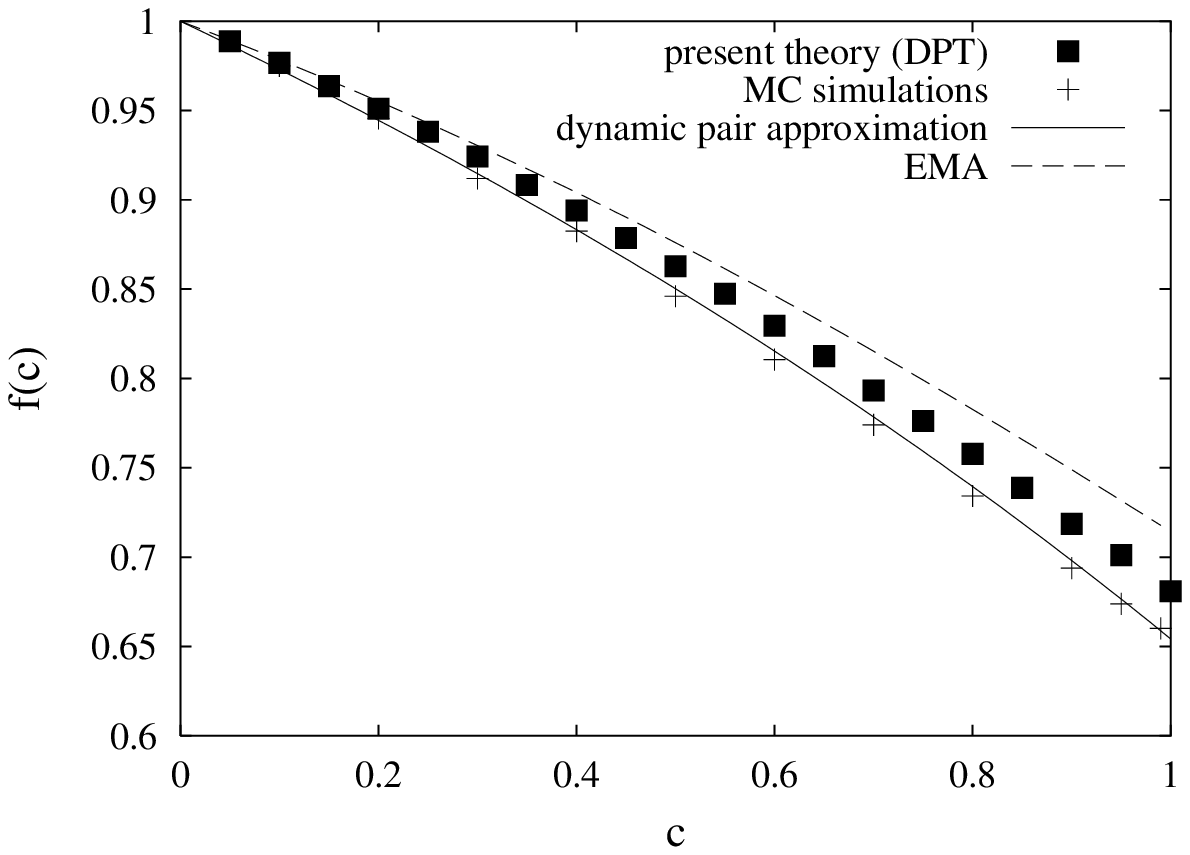,width=0.75\linewidth}
   Fig 2.
 \end{center}
\end{figure}
\begin{figure}[htb]
  \begin{center}
   \epsfig{file=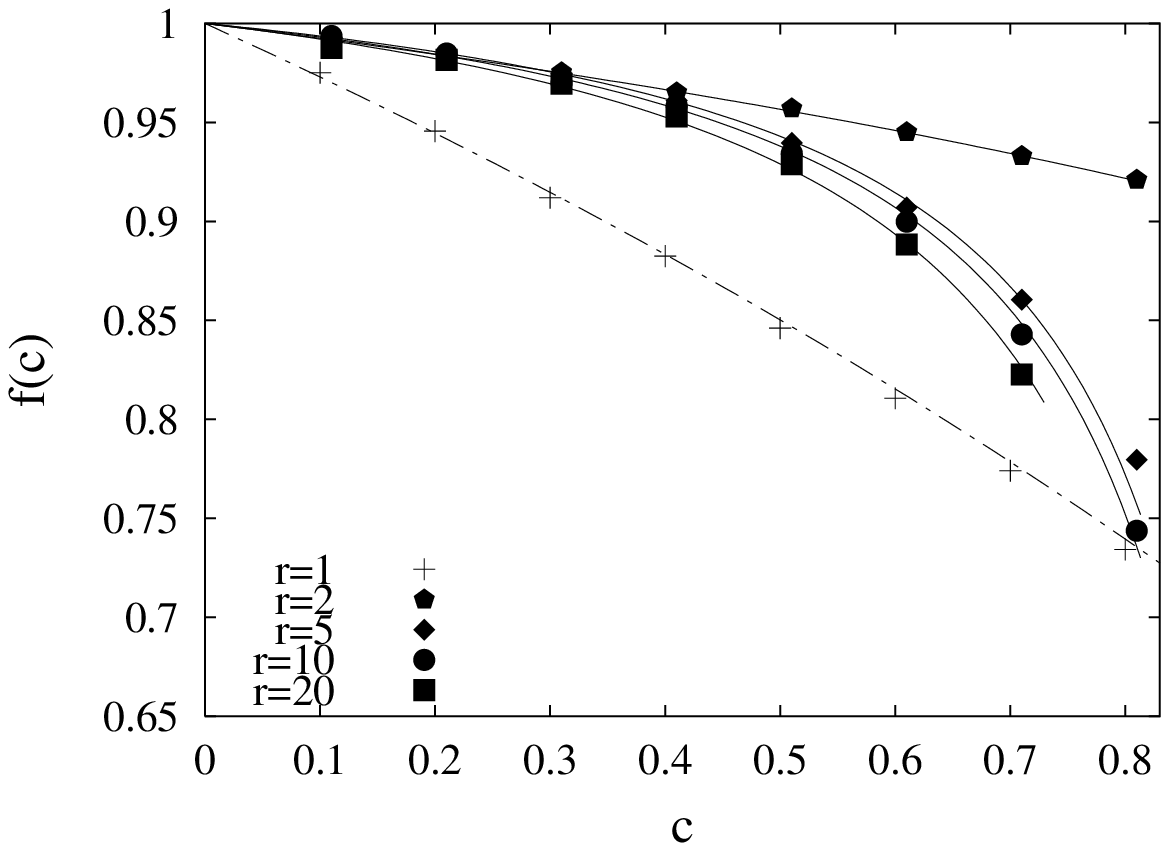,width=0.75\linewidth}
   Fig 3.
 \end{center}
\end{figure}
\begin{figure}[htb]
  \begin{center}
   \epsfig{file=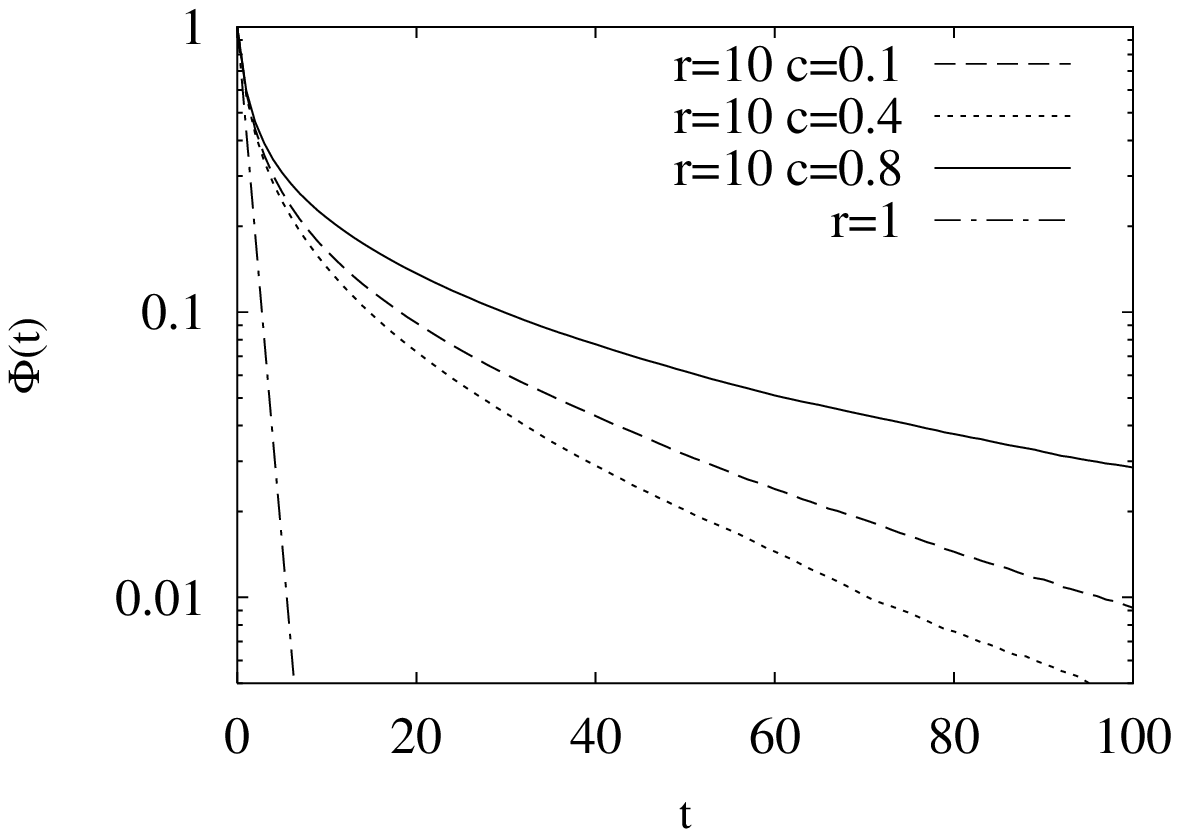,width=0.75\linewidth}
   Fig 4.
 \end{center}
\end{figure}
\begin{figure}[htb]
  \begin{center}
   \epsfig{file=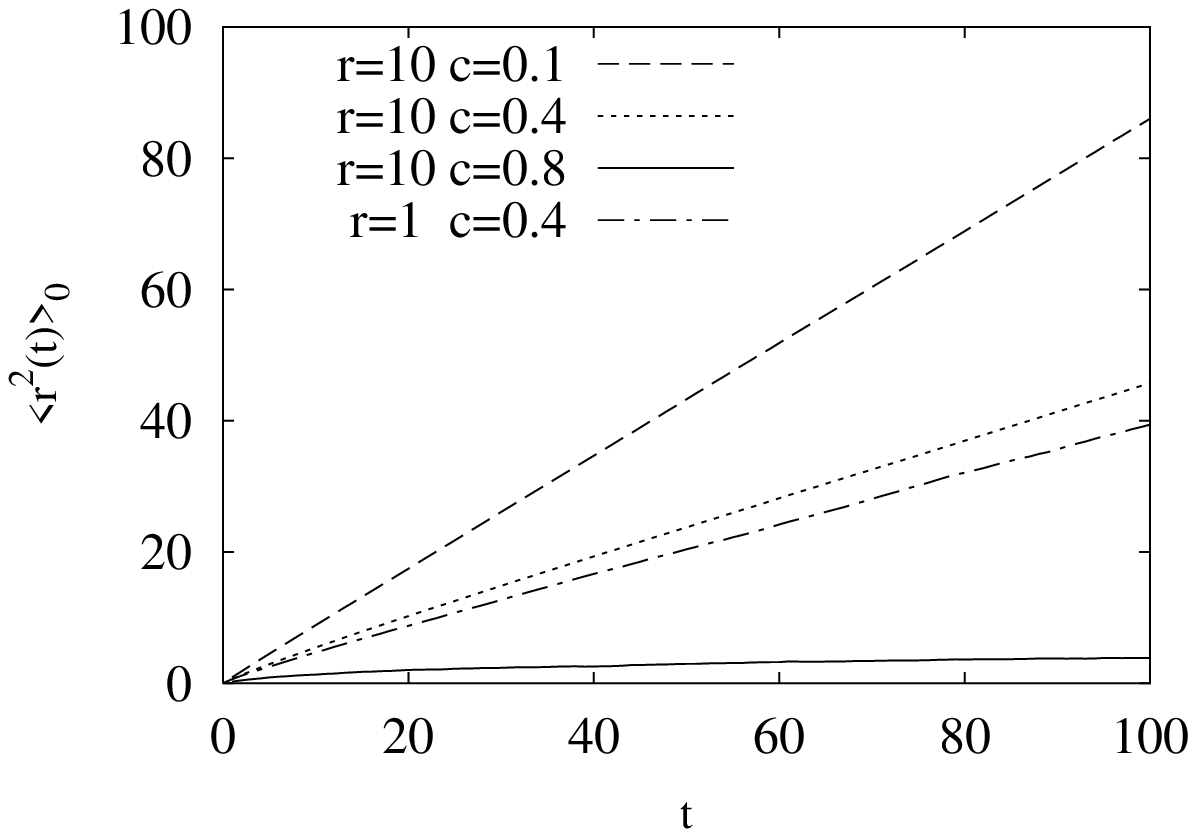,width=0.75\linewidth}
   Fig 5.
 \end{center}
\end{figure}
\begin{figure}[htb]
  \begin{center}
   \epsfig{file=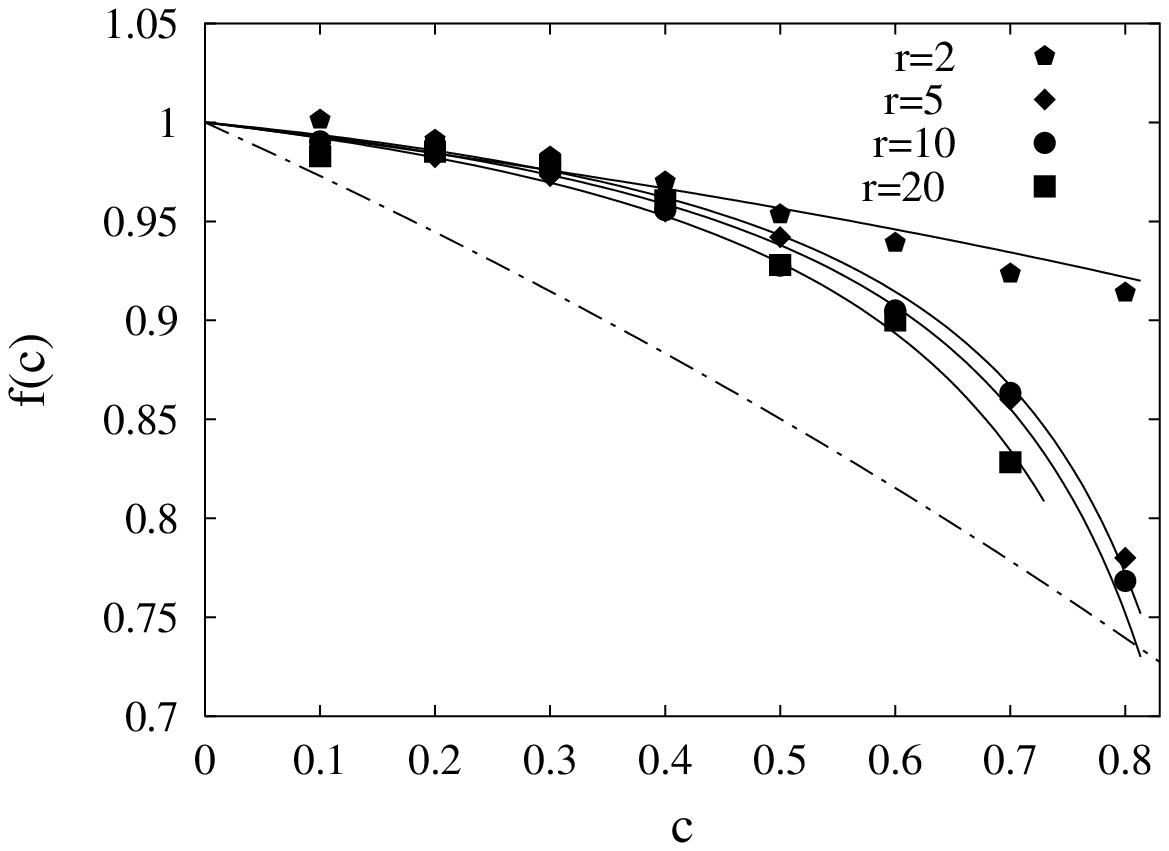,width=0.75\linewidth}
   Fig 6.
 \end{center}
\end{figure}
\begin{figure}[htb]
  \begin{center}
   \epsfig{file=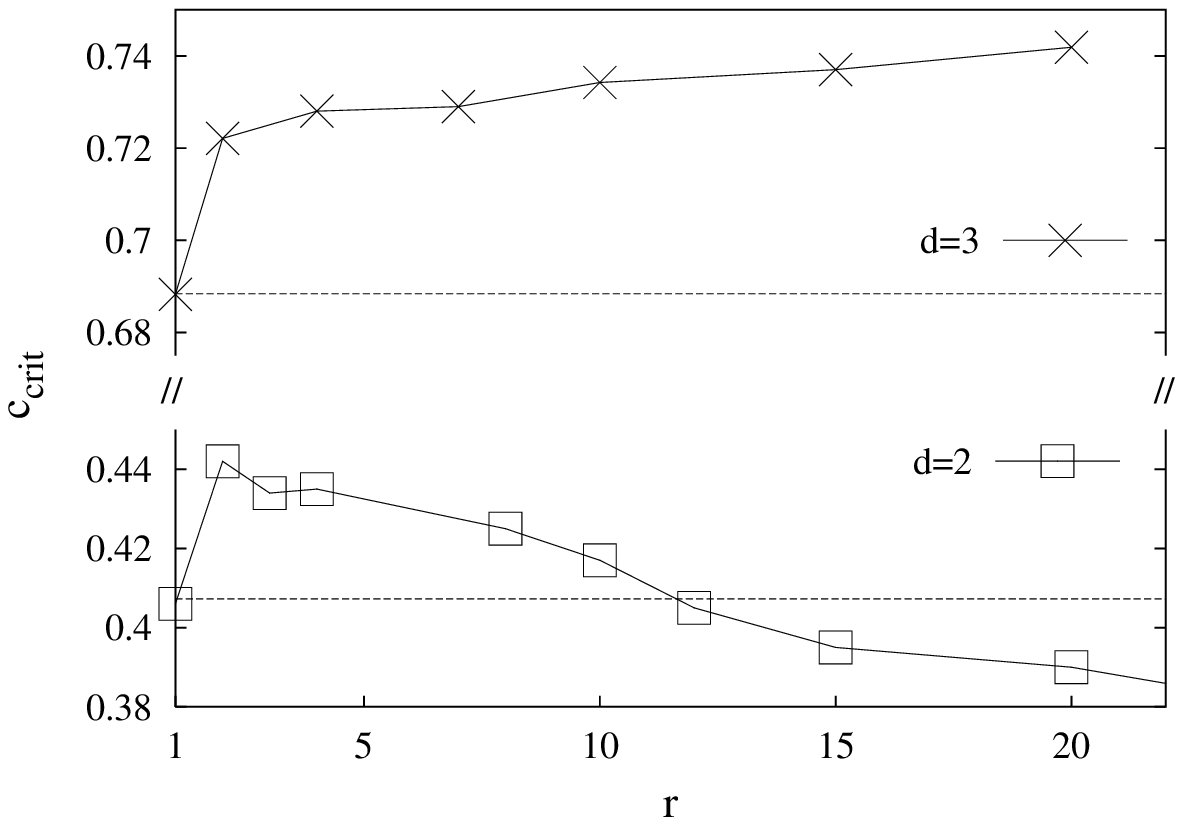,width=0.75\linewidth}
   Fig 7.
 \end{center}
\end{figure}

\begin{thebibliography}{ccccc}
\bibitem{15} In all parts of our present study we assume that hopping
  rates between two sites are symmetrical. Under this condition, the
  mean--square displacements with no renewal in the time--interval
  $[0,t]$, but in the two cases i) of the foregoing renewal taking
  place at $t_{0}=0$ and ii) of the foregoing renewal taking place at
  an arbitrary time $t_{0}<0$, are identical, as shown in
  Ref. \cite{Dru88}. In what follows we denote by $\langle
  (r(t))^{2}\rangle_{0}$ the mean--square displacement without any
  renewals under equilibrium conditions for the point particles.
\end{thebibliography}
\end{document}